\documentclass[aps,twocolumn,showpacs]{revtex4} 
\usepackage{latexsym}

\newcommand{\stac}[2]{\stackrel{\scriptscriptstyle {#1}}{#2}}

\begin{document}
%\draft

%<<<<<<<<<<<<< TITLE >>>>>>>>>>>>>>>%
\title{Asymmetric D-braneworld}

%<<<<<<<<<<<<< AUTHOR >>>>>>>>>>>>>>>%
\author{Keitaro Takahashi$^{(1)}$ and Tetsuya Shiromizu$^{(2,1,3)}$}

%<<<<<<<<<<<<< ADDRESS >>>>>>>>>>>>>>>%

\affiliation{$^{(1)}$Department of Physics, The University of Tokyo,  Tokyo 
113-0033, Japan}

\affiliation{$^{(2)}$Department of Physics, Tokyo Institute of Technology, 
Tokyo 152-8551, Japan}

\affiliation{$^{(3)}$Advanced Research Institute for Science and Engineering, 
Waseda University, Tokyo 169-8555, Japan}

%<<<<<<<<<<<<< DATE >>>>>>>>>>>>>>>%
\date{\today}

%======================================%
%<<<<<<<<<<<<< ABSTRACT >>>>>>>>>>>>>>>%
%======================================%
\begin{abstract}
In recent papers on Randall-Sundrum D-braneworld model with $Z_2$ symmetry,
it was shown that the effective gravity does not work as usual, that is,
the gravity does not couple to the gauge field localised on the brane in a usual way. 
At first glance there are two possibilities to avoid this serious problem. 
One is to remove the $Z_2$ symmetry and another is to consider a non-BPS state.
In this paper we analyze the Randall-Sundrum D-braneworld model without $Z_2$ symmetry
by long wave approximation. The result is unexpected one, that is, 
the gauge field does not couple to the gravity on the brane in the leading order again. 
Therefore the remaining possibility to recover the conventional gravitational theory 
would be non-BPS cases. 
\end{abstract}


\maketitle
%\vskip1cm

%======================================%
%<<<<<<<<<<<< SECTION I  >>>>>>>>>>>>>>%
%======================================%
%\baselineskip25pt
\label{sec:intro}
\section{Introduction}

New cosmological model, brane world, based on non-perturvative aspect of string theory
has been proposed few years ago \cite{Review}. The simplest model is Randall-Sundrum (RS)
model with warped compactification \cite{RSI,RSII}. So far it has been analysed intensely
in cosmological context because the self-gravity must be treated seriously and carefully.
As a consequence, the RS model and its extensions are basically able to reproduce the standard
cosmology. However, it is important to remember that the braneworld was inspired
by the D-brane, which has many interesting characteristics which the RS model does not have.
Therefore we want to ask whether a more realistic braneworld model based on 
D-brane works as well. Recently this issue was initiated in Ref. \cite{SKOT},
in which Born-Infeld action, bulk gauge fields and D-brane charge were appropriately
taking into account using the type IIB supergravity compactified on $S^5$.
See Refs. \cite{Dbrane,DBW1,DBW2,DBW3} for other issues based on D-brane.

A more tractable toy model has been investigated in \cite{OSKH,SHT,STH}.
There, the brane tension is assumed to equal to the brane charge and $Z_2$ symmetry
is imposed. Consequently, the gravity does not couple to the gauge fields at large distances
although the gauge field is supposed to localised on the brane. 
This is a serious problem if we want to use D-branes in a cosmological model.
A possible solution to this problem was discussed in Ref. \cite{SKT} assuming 
a non-BPS state, that is, a brane with a charge different from the tension.
As a result it was shown that the gauge field may couple to the gravity and
the gravitational constant is proportional to the cosmological constant on the brane. 

In this paper, we address another case, which has two D-branes and does not have $Z_2$ symmetry 
(See Ref. \cite{battye} for asymmetric braneworld model). It would be possible that
the $Z_2$ symmetry induces the irregular behavior obtained in Refs. \cite{SKOT,OSKH,SHT,STH}. 
Therefore we want to make clear the importance/unimportance of $Z_2$ symmetry. Surprisingly 
our conclusion obtained in the above papers is unchanged if the two form potentials for 
three form fields are continuous at the branes. To discuss this issue, we will employ
the gradient expansion method \cite{GE}. Recently it has been checked that such a method
can give us the same result as that obtained by the linear perturbation
at large distances \cite{STH}. 

The rest of this paper is organised as follows. In Sec. II, we describe the tractable toy model for 
D-branes. In Sec. III, we write down the field equations and the junction conditions. 
Then we solve the field equations under the junction conditions using the gradient 
expansion and derive the effective gravitational equation on the brane. In Sec. IV, we 
will give summary and discussion. In the appendix A, we sketch the conclusion 
obtained from the continuity of the two form field potentials.

%======================================%
%<<<<<<<<<<<< SECTION II  >>>>>>>>>>>>>%
%======================================%
%\baselineskip25pt
\section{Model}
\label{sec:model}

We consider the asymmetric Randall-Sundrum two brane model in type IIB supergravity compactified on 
$S^5$. The brane is described by Born-Infeld and Chern-Simons actions. So we begin with 
the following action(For example see Refs. \cite{SKOT,OSKH,SHT,STH,ST}.) 
%===========<Equation>============%
%
\begin{eqnarray}
S & = & \frac{1}{2\kappa^2} \int d^5x {\sqrt {-{\cal G}}}\biggl[{}^{(5)}R-2\Lambda 
-\frac{1}{2}|H|^2 
\nonumber \\ 
& & -\frac{1}{2}(\nabla \chi)^2-\frac{1}{2}|\tilde F|^2-\frac{1}{2}|\tilde 
G|^2 \biggr] \nonumber \\ 
& & +S_{\rm brane}^{(+)}+S_{\rm CS}^{(+)}+S_{\rm brane}^{(-)}+S_{\rm CS}^{(-)} , 
\label{action} 
\end{eqnarray} 
%
%=================================%
where $H_{MNK}=\frac{1}{2}\partial_{[M}B_{NK]}$, 
$F_{MNK}=\frac{1}{2}\partial_{[M}C_{NK]}$, 
$G_{K_1 K_2 K_3 K_4 K_5}=\frac{1}{4!}\partial_{[K_1}D_{K_2 K_3 K_4 K_5]}$, 
$\tilde F = F + \chi H$ and $\tilde G=G+C \wedge H$. $M,N,K=0,1,2,3,4$. 
$B_{MN}$ and $C_{MN}$ are 2-form fields, and $D_{K_1 K_2 K_3 K_4}$ is
a 4-form field. $\chi$ is a scalar field. ${\cal G}_{MN}$ is the metric 
of five dimensional spacetime. 

$S_{\rm brane}^{(\pm)}$ is given by Born-Infeld action
%===========<Equation>============%
%
\begin{eqnarray}
S_{\rm brane}^{(+)}=-\sigma \int d^4x {\sqrt {-{\rm det}(h+{\cal F}^{(+)})}}, 
\end{eqnarray}
%
%=================================%
%===========<Equation>============%
%
\begin{eqnarray}
S_{\rm brane}^{(-)}= \sigma \int d^4x {\sqrt {-{\rm det}(q+{\cal F}^{(-)})}}, 
\end{eqnarray} 
%
%=================================%
where $h_{\mu\nu}$ and $q_{\mu\nu}$ are the induced metric on the $D_{\pm}$-brane and 
%===========<Equation>============%
%
\begin{eqnarray}
{\cal F}_{\mu\nu}^{(\pm)}=B_{\mu\nu}^{(\pm)}+ \sigma^{-1/2}F_{\mu\nu}^{(\pm)}.
\end{eqnarray}
%
%=================================%
$F_{\mu\nu}$ is the $U(1)$ gauge field on the brane. Here $\mu,\nu=0,1,2,3$ and 
$\pm \sigma$ are $D_{\pm}$-brane tension. Hereafter $\sigma >0$ and then $D_-$-brane has 
the negative tension. 

$S_{\rm CS}^{(\pm)}$ is Chern-Simons action 
%===========<Equation>============%
%
\begin{eqnarray}
S_{\rm CS}^{(+)} & = & -\sigma \int d^4x {\sqrt {-h}} 
\epsilon^{\mu\nu\rho\sigma}\biggl[ \frac{1}{4}{\cal 
F}_{\mu\nu}^{(+)}C_{\rho\sigma}^{(+)}+\frac{\chi}{8}{\cal F}_{\mu\nu}^{(+)}{\cal F}_{\rho\sigma}^{(+)}
\nonumber \\
& & +\frac{1}{24}D_{\mu\nu\rho\sigma}^{(+)} \biggr], 
\end{eqnarray}
%
%=================================%
%===========<Equation>============%
%
\begin{eqnarray}
S_{\rm CS}^{(-)} & = & \sigma \int d^4x {\sqrt {-q}} 
\epsilon^{\mu\nu\rho\sigma}\biggl[ \frac{1}{4}{\cal 
F}_{\mu\nu}^{(-)}C_{\rho\sigma}^{(-)}+\frac{\chi}{8}{\cal F}_{\mu\nu}^{(-)}{\cal 
F}_{\rho\sigma}^{(-)} 
\nonumber \\
& & +\frac{1}{24}D_{\mu\nu\rho\sigma}^{(-)} \biggr], 
\end{eqnarray}
%
%=================================%
Here the brane charges are set equal to the brane tensions. Therefore, our model 
contains BPS state of D-branes.

%=======================================%
%<<<<<<<<<<<< SECTION III  >>>>>>>>>>>>>%
%=======================================%
\section{Basic equations}

In this section we write down the basic equations and boundary conditions. 
Let us perform (1+4)-decomposition along extra dimension 
%===========<Equation>============%
%
\begin{eqnarray}
ds^2={\cal G}_{MN}dx^{M}dx^{N}=e^{2\phi (x)}dy^2+g_{\mu\nu}(y,x) dx^\mu dx^\nu,
\end{eqnarray}
%
%=================================%
where $y$ is the coordinate orthogonal to the brane. 
$D_+$-brane and $D_-$-brane are supposed to locate at $y=y^{(+)}=0$ and $y=y^{(-)}=y_0$. 
 
The spacelike ``evolutional" equations to the $y$-direction are 
%===========<Equation>============%
%
\begin{eqnarray}
e^{-\phi} \partial_y K 
& = & {}^{(4)} R-\kappa^2 \biggl( {}^{(5)}T^\mu_\mu -\frac{4}{3}{}^{(5)}T^M_M \biggr) -K^2 
\nonumber \\
& & -e^{-\phi}D^2 e^\phi, 
\label{evoK}
\end{eqnarray}
%
%=================================%
%===========<Equation>============%
%
\begin{eqnarray}
e^{-\phi} \partial_y \tilde K^\mu_\nu & = &  {}^{(4)}\tilde R^\mu_\nu 
-\kappa^2\biggl({}^{(5)}T^\mu_\nu 
-\frac{1}{4} 
\delta^\mu_\nu {}^{(5)}T^\alpha_\alpha \biggr)-K \tilde K^\mu_\nu \nonumber \\ 
& & ~~-e^{-\phi}[D^\mu D_\nu e^{\phi}]_{\rm traceless}, 
\label{traceless} 
\end{eqnarray}
%
%=================================%
%===========<Equation>============%
%
\begin{eqnarray} 
\partial_y^2 \chi +D^2 \chi +e^\phi K\partial_y \chi-\frac{1}{2}H_{y\alpha\beta}\tilde 
F^{y\alpha\beta}=0, 
\end{eqnarray} 
%
%=================================%
%===========<Equation>============%
%
\begin{eqnarray}
& & \partial_y X^{y\mu\nu}+e^\phi KX^{y\mu\nu}+D_\alpha \phi H^{\alpha\mu\nu}+D_\alpha H^{\alpha\mu\nu} 
\nonumber \\ 
& & ~~~~+\frac{1}{2}F_{y\alpha\beta}\tilde G^{y\alpha\beta\mu\nu}=0, 
\label{evoH} 
\end{eqnarray} 
%
%=================================%
%===========<Equation>============%
%
\begin{eqnarray}
& & \partial_y \tilde F^{y\mu\nu}+e^\phi K \tilde F^{y\mu\nu}+D_\alpha \phi \tilde F^{\alpha\mu\nu}
+D_\alpha \tilde F^{\alpha\mu\nu}
\nonumber \\
& & ~~~~-\frac{1}{2}H_{y\alpha\beta}
\tilde G^{y\alpha\beta\mu\nu}=0,
\label{evoF}
\end{eqnarray}
%
%=================================%
%===========<Equation>============%
%
\begin{eqnarray}
\partial_y \tilde G_{y \alpha_1 \alpha_2 \alpha_3 \alpha_4}
=e^\phi K\tilde  G_{y \alpha_1 \alpha_2 \alpha_3 \alpha_4},
\end{eqnarray}
%
%=================================%
where $X^{y\mu\nu}:=H^{y\mu\nu}+\chi \tilde F^{y\mu\nu}$ and the 
energy-momentum tensor is 
%===========<Equation>============%
%
\begin{eqnarray}
&& \kappa^2\;{}^{(5)\!}T_{MN} =  \frac{1}{2}\biggl[ \nabla_M \chi \nabla_N \chi
-\frac{1}{2}g_{MN} (\nabla \chi)^2 \biggr]
\nonumber \\
& & ~~~~~~~~~~
+\frac{1}{4}\biggl[H_{MKL}H_N^{~KL}-g_{MN}|H|^2 \biggr] 
\nonumber \\
& & ~~~~~~~~~~
 +\frac{1}{4}\biggl[\tilde F_{MKL}\tilde
F_N^{~KL}-g_{MN}|\tilde F|^2
\biggr]
\nonumber \\
& & ~~~~~~~~~~
 +\frac{1}{96}\tilde G_{MK_1 K_2 K_3 K_4} \tilde G_{N}^{~~K_1
K_2 K_3 K_4}-\Lambda g_{MN}.
\nonumber \\
& & 
\end{eqnarray}
%
%=================================%
$K_{\mu\nu}$ is the extrinsic curvature, $K_{\mu\nu}=\frac{1}{2}e^{-\phi} \partial_y g_{\mu\nu}$. 
$\tilde K^\mu_\nu$ and ${}^{(4)}\tilde R^\mu_\nu$ are the traceless parts 
of $K^\mu_\nu$ and ${}^{(4)}R^\mu_\nu$, respectively. 
Here $D_\mu$ is the covariant derivative with respect to $g_{\mu\nu}$.

The constraints on $y={\rm const.}$ hypersurfaces are 
%===========<Equation>============%
%
\begin{eqnarray}
& & -\frac{1}{2}\biggl[{}^{(4)}R-\frac{3}{4}K^2+\tilde K^\mu_\nu \tilde K^\nu_\mu \biggr]
=\kappa^2\:{}^{(5)\!}T_{yy}e^{-2\phi}, 
\label{conK}
\end{eqnarray}
%
%=================================%
%===========<Equation>============%
%
\begin{eqnarray}
D_\nu K^\nu_\mu-D_\mu K = \kappa^2\:{}^{(5)\!}T_{\mu y}e^{-\phi},
\end{eqnarray}
%
%=================================%
%===========<Equation>============%
%
\begin{eqnarray}
D_\alpha(e^{\phi} X^{y\alpha\mu})+\frac{1}{6}e^\phi F_{\alpha_1 \alpha_2 \alpha_3} 
\tilde G^{y \alpha_1 \alpha_2 \alpha_3 \mu}= 0, \label{con1}
\end{eqnarray}
%
%=================================%
%===========<Equation>============%
%
\begin{eqnarray}
D_\alpha (e^{\phi} \tilde F^{y\alpha\mu})-\frac{1}{6}e^\phi H_{\alpha_1 \alpha_2 \alpha_3}
\tilde G^{y \alpha_1 \alpha_2 \alpha_3 \mu}=0, \label{con2}
\end{eqnarray}
%
%=================================%
%===========<Equation>============%
%
\begin{eqnarray}
D^\alpha (e^{-\phi} \tilde G_{y \alpha \mu_1 \mu_2 \mu_3})=0.
\end{eqnarray}
%
%=================================%

The junction conditions at the brane located $y=y^{(\pm)}$ are 
%===========<Equation>============%
%
\begin{eqnarray}
& & \lbrace K_{\mu\nu} - g_{\mu \nu} K \rbrace_{y=y^{(\pm)}}^{-}  = 
\kappa^2 \sigma (g_{\mu\nu}-T^{(\pm)}_{\mu\nu} ) \nonumber \\
& & ~~~~~~~~~~~~~~~~~~~~~~~~~~~~~~~~~~~~~~~~~+O(T_{\mu\nu}^2) \label{omit} \\ 
& & \lbrace H_{y\mu\nu}(y^{(\pm)},x) \rbrace^- = 2 \kappa^2 \sigma e^\phi {\cal F}_{\mu\nu}^{(\pm)}, \\ 
& & \lbrace \tilde F_{y\mu\nu}(y^{(\pm)},x) \rbrace^- 
=\kappa^2\sigma e^\phi \epsilon_{\mu\nu\alpha\beta}{\cal F}^{(\pm)\alpha\beta}, \\ 
& & \lbrace \tilde G_{y\mu\nu\alpha\beta}(y^{(\pm)},x) \rbrace^- 
=2 \kappa^2 \sigma e^\phi \epsilon_{\mu\nu\alpha\beta},\\ 
& & \lbrace \partial_y \chi (y^{(\pm)},x) \rbrace^- 
= \frac{\kappa^2}{4}\sigma e^\phi \epsilon^{\mu\nu\alpha\beta}{\cal F}^{(\pm)}_{\mu\nu}{\cal 
F}_{\alpha\beta}^{(\pm)}, 
\end{eqnarray}
%
%=================================%
where for any tensor field $Q$, $\lbrace Q \rbrace^-$ is defined as
$\lbrace Q \rbrace^- \equiv Q_R - Q_L$. Subscripts $R$ and $L$ denote the quantity evaluated
on the right and the left side of the $D_+$ brane, respectively. In the above 
%===========<Equation>============%
%
\begin{eqnarray}
T^{(\pm)\mu}_{~~~~~\nu}={\cal F}^{(\pm)\mu\alpha}{\cal F}^{(\pm)}_{\nu \alpha}
- \frac{1}{4}\delta^\mu_\nu {\cal F}_{\alpha\beta}^{(\pm)} {\cal F}^{(\pm) \alpha\beta}.
\end{eqnarray}
%
%=================================%

From the junction condition for $\chi$, we can omit the contribution of $\chi$ to the gravitational 
equation on the brane in the approximations which we will employ. Moreover, we omit the quadratic 
term in Eq. (\ref{omit}).

%======================================%
%<<<<<<<<<<<< SECTION IV  >>>>>>>>>>>>>%
%======================================%
\section{Effective theory}

In this section, we approximately solve the bulk field equations by long wave 
approximation (gradient expansion \cite{GE}) and derive the effective gravitational theory on the 
brane. 

In the case with bulk fields we must
carefully use the geometrical projection method \cite{battye,SMS} because the projected Weyl
tensor $E_{\mu\nu}$ contains the leading effect from the bulk fields. 

The bulk metric is written again as,
%===========<Equation>============%
%
\begin{eqnarray}
ds^2=e^{2\phi (x)}dy^2+g_{\mu\nu}(y,x) dx^\mu dx^\nu. 
\end{eqnarray}
%
%=================================%
The induced metric on the brane will be denoted by 
$h_{\mu\nu}:=g_{\mu\nu}(0,x)$ 
and then
%===========<Equation>============%
%
\begin{eqnarray}
g_{\mu\nu}(y,x)
=a^2(y,x)\Bigl[h_{\mu\nu}(x)+\stac{(1)}{g}_{\mu\nu}(y,x)+\cdots\Bigr].
\end{eqnarray}
%
%=================================%
In the above $\stac{(1)}{g}_{\mu\nu}(0,x)=0 $ and $a(0,x)=1$. In a similar way, 
the extrinsic curvature is expanded as 
%===========<Equation>============%
%
\begin{eqnarray}
K^\mu_\nu = \stac{(0)}{K^\mu_\nu}+ \stac{(1)}{K^\mu_\nu}+\stac{(2)}{K^\mu_\nu}+ 
\cdots.
\end{eqnarray}
%
%=================================%
The small parameter is $\epsilon = (\ell /L)^2 \ll 1$,
where $L$ and $\ell$ are the curvature scale on the brane 
and the bulk anti-deSitter curvature scale, respectively. 

%--------------------------------------%
%<<<<<<<<<< Subsection A  >>>>>>>>>>>>>%
%--------------------------------------%
\subsection{Background}

Without derivation we present the background spacetime. It is locally five dimensional 
anti-deSitter like spacetime
%===========<Equation>============%
%
\begin{eqnarray}
ds^2=e^{2\phi(x)}dy^2+a^2(y,x)h_{\mu\nu}dx^\mu dx^\nu 
\end{eqnarray}
%
%=================================%
where 
%===========<Equation>============%
%
\begin{eqnarray}
a=a_R = e^{-\frac{y}{\ell_R}e^\phi}~~{\rm for}~~0 \leq y \leq y_0=y^{(-)}
\end{eqnarray}
%
%=================================%
and
%===========<Equation>============%
%
\begin{eqnarray}
& & a=a_L = e^{\bigl[\frac{y}{\ell_L}-\bigl(\frac{1}{\ell_R}+\frac{1}{\ell_L} \bigr)y_0\bigr]e^\phi} \nonumber \\
& & ~~~~~~~~~~~~~~{\rm for}~~y_0 \leq y \leq y_*= \biggl(1+\frac{\ell_L}{\ell_R} \biggr)y_0.
\end{eqnarray}
%
%=================================%
The $y=y_*$ hypersurface is identified with the $y=0$ hypersurface. 

The junction condition for the extrinsic curvature are 
%===========<Equation>============%
%
\begin{eqnarray}
\stac{(0)}{K^R}_{\mu\nu}-\stac{(0)}{K^L}_{\mu\nu} = -\Biggl(\frac{1}{\ell_R}+\frac{1}{\ell_L} \Biggr) h_{\mu\nu}
=-\frac{\kappa^2}{3} \sigma h_{\mu\nu} \label{junex-}
\end{eqnarray}
%
%=================================%
and
%===========<Equation>============%
%
\begin{eqnarray}
\frac{\kappa^2}{2}\Biggl( \stac{(0)}{K^R}_{\mu\nu}+\stac{(0)}{K^L}_{\mu\nu} \Biggr)(-\sigma g^{\mu\nu}) & = &  -\lbrace {}^{(5)}G_{\mu\nu} 
n^\mu n^\nu \rbrace^- \nonumber \\
& = & -(\Lambda_R^{\rm tot} -\Lambda_L^{\rm tot}) \label{junex+}
\end{eqnarray}
%
%=================================%
where 
%===========<Equation>============%
%
\begin{eqnarray}
\Lambda^{\rm tot}_{R,L}=\Lambda_{R,L}-\frac{1}{96}
\tilde G^{R,L}_{y \alpha_1 \alpha_2 \alpha_3 \alpha_4} \tilde G_{R,L}^{y \alpha_1 \alpha_2 \alpha_3 \alpha_4}.
\end{eqnarray}
%
%=================================%
Eqs. (\ref{junex-}) and (\ref{junex+}) become 
%===========<Equation>============%
%
\begin{eqnarray}
\frac{1}{\ell_R}+\frac{1}{\ell_L}=\frac{\kappa^2}{3} \sigma
\end{eqnarray}
%
%=================================%
and
%===========<Equation>============%
%
\begin{eqnarray}
2\kappa^2 \sigma \Biggl(-\frac{1}{\ell_R}+\frac{1}{\ell_L}  \Biggr) =-\Lambda_R^{\rm tot} + \Lambda_L^{\rm tot}.
\end{eqnarray}
%
%=================================%

The junction condition for $\tilde G_5$ is 
%===========<Equation>============%
%
\begin{eqnarray}
\lbrace \tilde G_{y \alpha_1 \alpha_2 \alpha_3 \alpha_4}(y^{(\pm)},x) \rbrace^- 
=2\kappa^2 \sigma e^\phi \epsilon_{\alpha_1 \alpha_2 \alpha_3 \alpha_4}.
\end{eqnarray}
%
%=================================%
Since the solution are given by 
%===========<Equation>============%
%
\begin{eqnarray}
\tilde G^R_{y \alpha_1 \alpha_2 \alpha_3 \alpha_4}=\alpha_R a_R^4 e^\phi 
\epsilon_{\alpha_1 \alpha_2 \alpha_3 \alpha_4}
\end{eqnarray}
%
%=================================%
and
%===========<Equation>============%
%
\begin{eqnarray}
\tilde G^L_{y \alpha_1 \alpha_2 \alpha_3 \alpha_4}=\alpha_L a_L^4 e^\phi 
\epsilon_{\alpha_1 \alpha_2 \alpha_3 \alpha_4},
\end{eqnarray}
%
%=================================%
the junction condition becomes 
%===========<Equation>============%
%
\begin{eqnarray}
\lbrace \alpha \rbrace^- =2\kappa^2 \sigma. \label{junalpha}
\end{eqnarray}
%
%=================================%
Then the continuity condition for four form potential of $\tilde G_5 $ 
completely fix the value of $\alpha^R$ and $\alpha^L$. To see this we begin with 
%===========<Equation>============%
%
\begin{eqnarray}
\tilde G^{R,L}_{y \alpha_1 \alpha_2 \alpha_3 \alpha_4} = \partial_y D^{R,L}_{\alpha_1 \alpha_2 \alpha_3 \alpha_4}.
\end{eqnarray}
%
%=================================%
Then the potential $D^{R,L}_{\alpha_1 \alpha_2 \alpha_3 \alpha_4}$ can be solved as 
%===========<Equation>============%
%
\begin{eqnarray}
D^{R,L}_{\alpha_1 \alpha_2 \alpha_3 \alpha_4}= \mp \ell_{R,L}a^4_{R,L}
\epsilon_{\alpha_1 \alpha_2 \alpha_3 \alpha_4}+ d^{R,L}_{\alpha_1 \alpha_2 \alpha_3 \alpha_4}(x),
\end{eqnarray}
%
%=================================%
where $d^{R,L}_{\alpha_1 \alpha_2 \alpha_3 \alpha_4}(x)$ is the constant of integration. 
Here it is natural to assume the continuity $D^{R,L}_{\alpha_1 \alpha_2 \alpha_3 \alpha_4}$ on branes, 
that is, 
%===========<Equation>============%
%
\begin{eqnarray}
D^R_{\alpha_1 \alpha_2 \alpha_3 \alpha_4}(y^{(\pm)},x)= D^L_{\alpha_1 \alpha_2 \alpha_3 \alpha_4}(y^{(\pm)},x). 
\end{eqnarray}
%
%=================================%
Then we obtain 
%===========<Equation>============%
%
\begin{eqnarray}
\lbrace d_{\alpha_1 \alpha_2 \alpha_3 \alpha_4} \rbrace^- = \frac{1}{4}(\ell_L \alpha_L + \ell_R \alpha_R)
\epsilon_{\alpha_1 \alpha_2 \alpha_3 \alpha_4}
\end{eqnarray}
%
%=================================%
and
%===========<Equation>============%
%
\begin{eqnarray}
\lbrace d_{\alpha_1 \alpha_2 \alpha_3 \alpha_4} \rbrace^- = \frac{1}{4}(\ell_L \alpha_L + \ell_R \alpha_R)a_0^4
\epsilon_{\alpha_1 \alpha_2 \alpha_3 \alpha_4},
\end{eqnarray}
%
%=================================%
where $a_0=a_R(y=y_0)=e^{-\frac{y_0}{\ell_R} e^\phi}$. The above equations give us 
%===========<Equation>============%
%
\begin{eqnarray}
\frac{\alpha_R}{\alpha_L}=\frac{\frac{1}{\ell_R}}{-\frac{1}{\ell_L}}. 
\end{eqnarray}
%
%=================================%
Together with Eq. (\ref{junalpha}), then, we are resulted in 
%===========<Equation>============%
%
\begin{eqnarray}
\alpha_R = \frac{6}{\ell_R}~~{\rm and}~~\alpha_L=-\frac{6}{\ell_L}.
\end{eqnarray}
%
%=================================%
Hereafter we set RS tuning
%===========<Equation>============%
%
\begin{eqnarray}
\Lambda_{R,L}^{\rm tot}+\frac{6}{\ell^2_{R,L}}= \Lambda_{R,L}+\frac{15}{\ell_{R,L}^2}=0.
\end{eqnarray}
%
%=================================%
It means that the net cosmological constant on the brane vanishes.

\subsection{Form fields}

Let us focus on $H_3$ and $\tilde F_3$ fields which can contribute to the gravity on the brane.  
From first order differential equations for them we obtain the following second order
differential equation
%===========<Equation>============%
%
\begin{eqnarray}
\partial_y^2 H^{R,L}_{y\mu\nu} -\frac{36}{\ell_{R,L}^2}e^{2\phi} H^{R,L}_{y \mu \nu} =0.
\end{eqnarray}
%
%=================================%
The solution is given by 
%===========<Equation>============%
%
\begin{eqnarray}
H^{R,L}_{y\mu\nu} = a^{-6}_{R,L}\alpha^{R,L}_{\mu\nu}+a^{6}_{R,L} \beta^{R,L}_{\mu\nu}
\end{eqnarray}
%
%=================================%
and
%===========<Equation>============%
%
\begin{eqnarray}
\tilde F^{R,L}_{y\mu\nu} & = & \pm \frac{\ell_{R,L}}{12} \epsilon_{\mu\nu}^{~~~\alpha\beta} \partial_y 
H_{y \alpha \beta}^{R,L} \nonumber \\
& = & \frac{1}{2}\epsilon_{\mu\nu}^{~~~\alpha\beta}(a_{R,L}^{-6}\alpha^{R,L}_{\mu\nu}
-a^6_{R,L} \beta^{R,L}_{\mu\nu}).
\end{eqnarray}
%
%=================================%
The junction condition at $y=0$ implies
%===========<Equation>============%
%
\begin{eqnarray}
\lbrace \alpha_{\mu\nu} \rbrace^- = 2\kappa^2 \sigma e^\phi {\cal F}_{\mu\nu}^{(+)} \label{junF0}
\end{eqnarray}
%
%=================================%
and
%===========<Equation>============%
%
\begin{eqnarray}
\lbrace \beta_{\mu\nu} \rbrace^- = 0. \label{junbeta0}
\end{eqnarray}
%
%=================================%
In the same way, the junction conditions on $y=y_0$ provides us 
%===========<Equation>============%
%
\begin{eqnarray}
a_0^{-6} \lbrace \alpha_{\mu\nu} \rbrace^- = 2\kappa^2 \sigma e^\phi {\cal F}_{\mu\nu}^{(-)} \label{junFy0}
\end{eqnarray}
%
%=================================%
and
%===========<Equation>============%
%
\begin{eqnarray}
\lbrace \beta_{\mu\nu} \rbrace^- = 0. 
\end{eqnarray}
%
%=================================%
Then Eqs. (\ref{junF0}) and (\ref{junFy0}) lead us 
%===========<Equation>============%
%
\begin{eqnarray}
{\cal F}^{(+)}_{\mu \nu} = a^6_0  {\cal F}^{(-)}_{\mu\nu}. \label{relation}
\end{eqnarray}
%
%=================================%
Finally the continuity for the potential $B_2$ and $C_2$ of $H_3$ and $F_3$ 
determine $\alpha_{\mu\nu}^{R,L}$ and $\beta_{\mu\nu}^{R,L}$ as 
%===========<Equation>============%
%
\begin{eqnarray}
\alpha_{\mu\nu}^{R,L}=\pm \frac{6}{\ell_{R,L}}e^\phi {\cal F}^{(+)}_{\mu\nu}
\end{eqnarray}
%
%=================================%
and 
%===========<Equation>============%
%
\begin{eqnarray}
\beta_{\mu\nu}^{R,L}=0. 
\end{eqnarray}
%
%=================================%
See the appendix A for the detail of the argument on continuity of $B_2$. 

As a consequence, we can uniquely determine $H_{y \mu\nu}$ and $\tilde F_{y \mu\nu}$
%===========<Equation>============%
%
\begin{eqnarray}
H_{y \mu\nu}^{R,L}= \pm \frac{6}{\ell_{R,L}} e^\phi a^{-6}_{R,L}{\cal F}^{(+)}_{\mu\nu}
\end{eqnarray}
%
%=================================%
and
%===========<Equation>============%
%
\begin{eqnarray}
\tilde F_{y \mu\nu}^{R,L}= \pm \frac{3}{\ell_{R,L}} e^\phi a^{-6}_{R,L}\epsilon_{\mu\nu}^{~~~\alpha\beta} 
{\cal F}^{(+)}_{\alpha \beta}.
\end{eqnarray}
%
%=================================%

\subsection{Extrinsic curvature and effective theory}

It is now ready to derive the effective theory on the brane. To do so we will 
solve the evolutional equation of extrinsic curvature. The solution of the traceless part is 
%===========<Equation>============%
%
\begin{eqnarray}
\stac{(1)}{\tilde K^{R\mu}_{~~\nu}} & = & -\frac{\ell_R}{2a_R^2} {}^{(4)} \tilde R^\mu_\nu
-\frac{3}{\ell_R} a^{-16}_R T^{(+) \mu }_{~~~~\nu} 
+\frac{\chi^{\mu R}_\nu}{a^4_R} \nonumber \\
& & -\frac{1}{a_R^2} \Biggl[{\cal D}^\mu {\cal D}_\nu d_R -\frac{1}{\ell_R}{\cal D}^\mu d_R 
{\cal D}_\nu d_R \Biggr]_{\rm traceless}
\label{solKR}
\end{eqnarray}
%
%=================================%
and
%===========<Equation>============%
%
\begin{eqnarray}
\stac{(1)}{\tilde K^{L \mu}_{~~\nu}} & = & \frac{\ell_L}{2a_L^2} {}^{(4)} \tilde R^\mu_\nu
+\frac{3}{\ell_L} a^{-16}_L T^{(+)\mu}_{~~~~\nu} +\frac{\chi^{\mu L}_\nu}{a^4_L}\nonumber \\
& & +\frac{1}{a_L^2}\Biggl[{\cal D}^\mu {\cal D}_\nu d_L -\frac{1}{\ell_L}{\cal D}^\mu d_L {\cal D}_\nu d_L 
\Biggr]_{\rm traceless}, \label{solKL}
\end{eqnarray}
%
%=================================%
where $\chi_{\mu\nu}^{R,L}$ are the constants of integration and $d \equiv y e^\phi$ is
the proper distance between the two D-branes, which is called radion.
${\cal D}_\mu$ is the covariant derivative with respect to $h_{\mu\nu}$. 
The junction condition for $\tilde K^\mu_\nu$ at $y=0$ and the above solution give us 
%===========<Equation>============%
%
\begin{eqnarray}
-\kappa^2 \sigma T^{ (+) \mu}_{~~~~~\nu} & =  & \lbrace \stac{(1)}{\tilde K^\mu_\nu} \rbrace^- \nonumber \\
& = & -\frac{1}{2}(\ell_R + \ell_L){}^{(4)}\tilde R^\mu_\nu \nonumber \\
& & -3 \Biggl( \frac{1}{\ell_R}+\frac{1}{\ell_L} \Biggr) T^{(+)\mu}_{~~~~\nu} 
+\lbrace \chi^\mu_\nu \rbrace^- .
\end{eqnarray}
%
%=================================%
The stress tensor in right-hand side is exactly canceled out with that in right-hand side! 
Thus the gravitational equation becomes 
%===========<Equation>============%
%
\begin{eqnarray}
{}^{(4)}G_{\mu\nu}(h)=\frac{2}{\ell_R + \ell_L} \lbrace \chi_{\mu\nu} \rbrace^- . \label{effective1}
\end{eqnarray}
%
%=================================%
$\lbrace \chi_{\mu\nu} \rbrace^-$ can be determined by the remaining junction condition 
at $y=y_0$. From the junction condition, indeed, we first obtain 
%===========<Equation>============%
%
\begin{eqnarray}
-2\kappa^2 \sigma T^{(-) \mu}_{~~~~\nu}
& = & -\frac{1}{2}a_0^{-2} (\ell_R + \ell_L){}^{(4)}\tilde R^\mu_\nu \nonumber \\
& & -3 \bigl(\frac{1}{\ell_R}+\frac{1}{\ell_L} \bigr)a_0^{-16}T^{(+)\mu}_{~~~~\nu}
    + a_0^{-4} \lbrace \chi^\mu_\nu \rbrace^- \nonumber \\
& & -a^{-2}_0 \Biggl(1+\frac{\ell_L}{\ell_R} \Biggr)
    \Biggl[{\cal D}^\mu {\cal D}_\nu d_R^0 \nonumber \\
& & -\frac{1}{\ell_R} {\cal D}^\mu d_R^0 {\cal D}_\nu d_R^0  \Biggr]_{\rm traceless},
\label{effective2}
\end{eqnarray}
%
%=================================%
where $d_R^0 = d_R(y=y_0)$ and we used the fact $d_L^0 = (\ell_L /\ell_R) d_R^0$. 
In the same way with the argument at $y=0$ brane, using of Eq. (\ref{relation}), 
we can show the left-hand side is exactly canceled with the second term in right-hand side. 
Finally Eqs. (\ref{effective1}) and (\ref{effective2}) can be summarised as  
%===========<Equation>============%
%
\begin{eqnarray}
{}^{(4)}G_{\mu\nu}(h)
& = & \frac{2}{\ell_R (a_0^{-2}-1)}  \nonumber \\
& & \times \biggl[ {\cal D}_\mu {\cal D}_\nu d_R
                    - \frac{1}{\ell_R}{\cal D}_\mu d_R {\cal D}_\nu d_R 
           \biggr]_{\rm traceless}.
\end{eqnarray}
%
%=================================%
This is our main result. 

The equation for radion $d_R^0$ can be derived from the trace part of the extrinsic curvature
and then 
%===========<Equation>============%
%
\begin{eqnarray}
{\cal D}^2d_R -\frac{1}{\ell_R}({\cal D}d_R)^2=0.
\end{eqnarray}
%
%=================================%

%======================================%
%<<<<<<<<<<<<< SECTION V  >>>>>>>>>>>>>%
%======================================%
%\baselineskip25pt
\section{Summary and discussion}
\label{sec:summary}

In this paper we investigated D-braneworld model without $Z_2$ symmetry and 
derived the effective theory on the D-brane. Surprisingly the gauge fields do not 
couple to the gravity on the D-brane at large distances. The result is basically the same
as that in the previous works \cite{SKOT,OSKH,SHT,STH}. 

Thus, the remaining possibility to recover the conventional gravitational theory in
D-braneworld would be non-BPS cases. In non-BPS state, a non-zero cosmological constant
appears on the brane. In the one-D-brane model discussed Ref. \cite{SKT}, indeed,
the appearance of the gravitational coupling to the gauge field localised on the brane
was confirmed using the gradient expansion method. Moreover it turned out that
the gravitational constant is proportional to the cosmological constant on the brane.
Therefore the presence of the cosmological constant on the branes seems to be the only solution
to the D-braneworld model if $B_2$, $C_2$ and $D_4$ are continuous at the branes.

%======================================%
%<<<<<<<<< Acknowledgements  >>>>>>>>>>>%
%======================================%
%\baselineskip25pt

\section*{Acknowledgements}

TS thank N. Sakai for useful discussion. KT would like to thank Y. Himemoto for
discussion on braneworld without $Z_2$ symmetry.
The work of TS was supported by Grant-in-Aid for Scientific
Research from Ministry of Education, Science, Sports and Culture of 
Japan(No.13135208, No.14740155 and No.14102004). 
The works of KT was supported by JSPS.

\vskip 5mm

\appendix

%=====================================%
%<<<<<<<<<<< APPENDIX A  >>>>>>>>>>>>>%
%=====================================%
%\baselineskip25pt

\section{Continuity for $B_2$ and $C_2$}

The potential $B_2$ and $C_2$ have the following solution 
%===========<Equation>============%
%
\begin{eqnarray}
B_{\mu\nu}^{R,L} & = & \mp \frac{\ell_{R,L}}{6}(a^{-6}_{R,L}\alpha^{R,L}_{\mu\nu}(x)-a^6_{R,L}\beta^{R,L}_{\mu\nu}(x) )
\nonumber \\
& & +b^{R,L}_{\mu\nu}(x)
\end{eqnarray}
%
%=================================%
and
%===========<Equation>============%
%
\begin{eqnarray}
C_{\mu\nu}^{R,L} & = & \mp \frac{\ell_{R,L}}{12} \epsilon_{\mu\nu}^{~~~\alpha\beta}
(a^{-6}_{R,L}\alpha^{R,L}_{\alpha\beta}(x)+a^6_{R,L}\beta^{R,L}_{\alpha\beta}(x) ) \nonumber \\
& & +c^{R,L}_{\mu\nu}(x),
\end{eqnarray}
%
%=================================%
where $b^{R,L}_{\mu \nu}(x)$ and $c^{R,L}_{\mu \nu}(x)$ are 
the constant of integration. From the continuity at $y=y^{(\pm)}$ it is easy to obtain 
%===========<Equation>============%
%
\begin{eqnarray}
\ell_R \alpha^R_{\mu\nu}+ \ell_L \alpha^L_{\mu\nu}=0
\end{eqnarray}
%
%=================================%
and
%===========<Equation>============%
%
\begin{eqnarray}
\ell_R \beta^R_{\mu\nu}+ \ell_L \beta^L_{\mu\nu}=0.
\end{eqnarray}
%
%=================================%
Together with Eqs. (\ref{junF0}) and (\ref{junbeta0}) we see
%===========<Equation>============%
%
\begin{eqnarray}
\alpha_{\mu\nu}^{R,L}= \mp \frac{6}{\ell_{R,L}}e^\phi {\cal F}^{(+)}_{\mu\nu}
\end{eqnarray}
%
%=================================%
and
%===========<Equation>============%
%
\begin{eqnarray}
\beta^{R,L}_{\mu\nu}=0.
\end{eqnarray}
%
%=================================%
These result are what we wanted to show.


\begin{thebibliography}{22}


\bibitem{Review}
G. Gabadadze, {\it ICTP lectures on large extra dimensions}, hep-ph/0308112; 
R. Maartens, gr-qc/0312059;
P. Brax, C. van de Bruck and A. Davis, hep-th/0404011; 
C. Csaki, {\it TASI Lectures on extra dimensions and branes}, hep-ph/0404096. 

\bibitem{RSI}
L.~Randall and R.~Sundrum, Phys. Rev. Lett. {\bf 83}, 3370 (1999).

\bibitem{RSII}
L.~Randall and R.~Sundrum, Phys. Rev. Lett. {\bf 83}, 4690 (1999).

\bibitem{SKOT}
T. Shiromizu, K. Koyama, S. Onda and T. Torii, Phys. Rev. {\bf D68}, 063506(2003).

\bibitem{Dbrane}
S. Kachru, R. Kallosh, A. Linde, J. Maldacena, Liam McAllister and S. P. Trivedi, JCAP {\bf 0310},013(2003).

\bibitem{DBW1}
C. P. Burgess, P. Martineau, F. Quevedo and R. Rabadan, JHEP {\bf 06}, 037(2003);
C. P. Burgess, N. E. Grandi, F. Quevedo and R. Rabadan, JHEP {\bf 0401}, 067(2004);
K. Takahashi and K. Ichikawa, Phys. Rev. {\bf D69},103506(2004);
E. J. Copeland, R. C. Myers and J. Polchinski, JHEP {\bf 0406}, 013(2004).

\bibitem{DBW2}
T. Shiromizu, T. Torii and T. Uesugi, Phys. Rev. {\bf D67}, 123517(2003);
M. Sami, N. Dadhich and T. Shiromizu, Phys. Lett. {\bf B568},118(2003);
E. Elizalde, J. E. Lidsey, S. Nojiri and S. D. Odintsov, hep-th/0307177;
T. Uesugi, T. Shiromizu, T. Torii and K. Takahashi, Phys. Rev. {\bf D69}, 043511(2004). 

\bibitem{DBW3}
S. B. Giddings, S. Kachru and J. Polchinski, Phys. Rev. {\bf D66}, 106006(2002);
O. DeWolfe and S. B. Giddings, Phys. Rev. {\bf D67}, 066008(2002).

\bibitem{OSKH}
S. Onda, T. Shiromizu, K. Koyama and S. Hayakawa,  Phys. Rev. {\bf D69},123503(2004). 

\bibitem{SHT}
T. Shiromizu, Y. Himemoto and K. Takahashi, hep-th/0405071. 

\bibitem{STH}
T. Shiromizu, K. Takahashi, Y. Himemoto and S. Yamamoto, hep-th/0407268. 

\bibitem{SKT}
T. Shiromizu, K. Koyama and T. Torii, Phys. Rev. {\bf D68}, 103513(2003).  

\bibitem{battye}
R. A. Battye, B. Carter, A. Mennim and J. Uzan, Phys. Rev. {\bf D64}, 124007(2001).
N.Kaloper, Phys. Rev. {\bf D60}, 123506(1999);
B.Carter, J. Uzan, R. A. Battye and A. Mennim, Class. Quantum Grav. {\bf 18}, 4871(2001);
A. Davis, I. Vernon, S. C. Davis and W. B. Rekins, Phys. Let. {\bf B504}, 254(2001);
O. Castillo-Felisola, A. Melfo, N. Pantoja and A. Ramirez, hep-th/0404083;
A. Paddila, hep-th/0406157.

\bibitem{GE}
T. Wiseman, Class. Quant. Grav. {\bf 19}, 3083(2002);
S. Kanno and J. Soda, Phys. Rev. D{\bf 66}, 043526(2002);
{\rm ibid}, 083506,(2002);
T. Shiromizu and K. Koyama, Phys. Rev. D{\bf 67}, 084022(2003);
S. Kanno and J. Soda, Gen. Rel. Grav. {\bf 36}, 689(2004). 

\bibitem{ST}
M. Sato and A. Tsuchiya, Prog. Theor. Phys. {\bf 109}, 687(2003).

\bibitem{SMS}
T. Shiromizu, K. Maeda and M. Sasaki, Phys. Rev. {\bf D62}, 024012(2000).


\end{thebibliography}
\end{document}